\documentclass[a4paper,oneside]{saip}
\usepackage{helvet}          % to have better matching fonts between Latex and Word (Title/Headings)
\usepackage{times}           % to have better matching fonts between Latex and Word (bulk text)
\usepackage{courierten}   % to have better matching fonts between Latex and Word (monospaced)
\usepackage[T1]{fontenc} % needed to deal with these fonts
\usepackage{graphicx}  %to include figures
\usepackage{url}
\usepackage{geometry} %	need:
\begin{document}

\title{Search for persistent radio emission towards selected localised Fast Radio Burst positions using the MeerKAT Telescope}

\author{Thulo Letsele$^{1}$, Lebogang Mfulwane$^{1}$, Christo Venter$^{1,2}$, James O. Chibueze $^{3,4}$, and Mechiel Christiaan Bezuidenhout$^{5,1}$}

\affil{$^1$Centre for Space Research, North-West University, Private Bag X6001, Potchefstroom 2520, South Africa}
\affil{$^2$National Institute for Theoretical and Computational Sciences (NITheCS), Potchefstroom, 2520, South Africa
}
\affil{$^3$Department of Mathematical Sciences, University of South Africa, Cnr Christian de Wet Rd and Pioneer Avenue, Florida Park, 1709, Roodepoort, South Africa}
\affil{$^4$Department of Physics and Astronomy, Faculty of Physical Sciences,  University of Nigeria, Carver Building, 1 University Road, Nsukka 410001, Nigeria}
\affil{$^5$SKA Observatory, 2 Fir Street, Observatory 7925, Cape Town, South Africa}

%only put the email of the main corresponding author, not of all authors.
\email{\url{thuloletsele1999@gmail.com}}

\begin{abstract}
Fast Radio Bursts (FRBs) are millisecond-duration radio pulses originating from cosmological distances, as indicated by their large dispersion measures. While numerous FRBs have now been localised to their host galaxies, a distinct class of compact electromagnetic counterpart, a Persistent Radio Source (PRS), has also been identified in some cases. Currently, only three, and possible a fourth repeating FRBs (FRB20121102A, FRB20190417A, FRB20190520B, and FRB20240114A) have confirmed associations with a PRS. Insight into progenitors, local environments, and the evolution of FRBs can be clarified by characterising these PRSs. In this work, we present 2 detected candidate PRSs using MeerKAT radio telescope data and one non-detection (as part of a larger study involving 25 FRB positions). Both FRB20221106 and FRB20181112 were found to have a host galaxy, and whether the detected radio continuum emission comes from the host galaxy or PRS is still an open question. High-resolution observations from a telescope such as e-MERLIN are required to resolve this question. If a compact PRS is detected, this telescope will provide the size, and investigate the flux variability and spectral shape of this compact PRS. Lastly, in the case of FRB220190102, which was observed over two epochs, no radio continuum was detected. However, a flux upper limit is provided for both epochs.
\end{abstract}

\section{Introduction}
Radio transients with signals lasting in the range of micro to milliseconds, and characterised by luminosities that make them visible over extragalactic distances \cite{Petroff,Lorimer}, are known as Fast Radio Bursts (FRBs). They were first discovered in 2007 when Duncan Lorimer and his student were analysing archival data of Manchester et al. \cite{Manchester}, searching for a sub-population of pulsars known as rotating radio transients. FRBs were believed to be one-off events until the discovery of the first repeating FRB \cite{Spitler}, which was localised with the Karl G. Jansky Very Large Array (VLA) to a dwarf star-forming galaxy of redshift of $z=0.19$ following its discovery \cite{Chatterjee,Tendulkar}. The mechanism behind the radiation and the origin of FRBs remains a mystery.\\
\\
Many researchers have proposed different possible progenitors for FRBs. These include the merging of a white dwarf and a black hole \cite{Mingarelli2015}, the merger of binary white dwarf stars \cite{Kashiyama2013}, the collision of two neutron stars \cite{Totani2013}, and the interaction of neutron stars with active galactic nuclei \cite{Vieyro}\footnote{Authors provide one possible interaction scenario, which involves a neutron star or magnetar in the vicinity of a low-luminosity AGN. The relativistic electron--position beams from the AGN can impact the surrounding plasma cloud, which includes those associated with a neutron star environment (that is how they interact). This interaction can trigger shocks and bring about instabilities that generate coherent radio emission, which can provide a possible explanation for repeating FRBs in AGN environments.}. The magnetar model became an interesting model because \cite{Popov} suggested that magnetar flares can produce FRBs. In addition, the association of the Galactic magnetar SGR J1935+2154 with FRB20200428 provided strong support for the magnetar model \cite{Bochenek2020,Andersen}.\\
\\
The environment in which FRBs occur is also a subject of interest to researchers. A Persistent Radio Source (PRS) observed in the vicinity of the FRB environment may provide important constraints on this environment. Currently, there are three, and possible a fourth actively repeating FRBs associated with PRSs \cite{moroianu2025,Niu2022,Chatterjee,bruni2024}. Most of these PRSs show high Faraday rotation measures (RMs), which suggests that the environment contains a compact nebula that is highly magnetised. Bruni et al.~\cite{bruni202} suggested that low RMs cannot be detected, and this can explain the low number of FRBs with PRSs. The nature of the PRS also remains a mystery, and scenarios proposed to explain this include a connection between supernova remnants and FRBs \cite{Connor} and binary systems \cite{Zhang}.\\
\\
Three FRBs, namely FRB20181112, FRB20190102, and FRB20221106, form part of this work. FRB20181112 and FRB20190102 show dispersion measures (DMs) of 589.27 pc cm$^{-3}$ and 363.6 pc cm$^{-3}$, with corresponding redshifts of $z=0.4755$ and $z= 0.291$, respectively \cite{Heintz}. Both FRB20181112 and FRB20190102 are non-repeating FRBs that have significantly contributed to the Macquart relation and cosmological constraints \cite{Macquart}. Lastly, we include FRB20221106, which remains one of the least studied FRBs; not much information about this FRB is available.\\
\\
The goal of this work is to expand this list of FRBs with identified PRSs. This will be done by searching for persistent radio emissions at FRB positions that have been well localised by the Australian Square Kilometre Array Pathfinder (ASKAP) telescope. If a radio source is detected, multiwavelength follow-up observations have to be conducted to confirm the existence of a PRS. However, if a radio source is not detected, this work will provide the upper flux limit. This work is part of a larger project involving follow-up of 25 well-localised FRB positions \cite{Chibueze,Aharonian25, sislebo}.\\
The paper is divided as follows: in Section~\ref{Data}, we present MeerKAT data reduction and describe how the images are produced. In Section~\ref{results}, we provide our results, a subset of those presented elsewhere \cite{sislebo}. We conclude in Section~\ref{conclusion}.

\section{Observations and Imaging} \label{Data}
The observational campaigns were conducted with the MeerKAT radio telescope for open time proposals in 2021, 2022, and 2023 (SCI-20210212-CV-01, SCI-20220822-CV-01, SCI-20230907-CV-01, respectively). The observations were done using the L-band (856-1712 MHz) receiver, and the FRB positions were observed for 120 minutes on source for each epoch, with a phase calibrator observed for 2 minutes, every 15 minutes. Tables~1 and~2 contain important observation details, including observation date, Right Ascension (RA), Declination (Dec), and synthesised beam of the observed FRBs.
\subsection{Calibration and Imaging}
MeerKAT data were obtained in visibility format, and were converted into a Measurement Set (MS) format using the KAT Data Access Library$\footnote{\url{www.github.com/ska-sa/katdal}}$. The \textsc{oxkat3}\cite{2020Heywood} analysis pipeline flags the low-gain bandpass edges on all baselines. In addition, radio frequency interferences are also flagged, and in cases where they might affect data, the \texttt{CASA} task such as \texttt{RFLAG} is used. The \texttt{TRICOLOR} and \texttt{TFCROP} packages are used for the target field and calibrators, respectively \cite{2007Mcmullin}.\\
\\
The \textsc{oxkat3} pipeline uses tasks from \texttt{CASA} to obtain the flux scale and corrections for residual delay calibrators, time-varying gain, and bandpass in the case of cross-calibration \cite{2007Mcmullin}. The corrections obtained are applied to the target field, and calibrated visibilities of the target field are extracted to obtain the science target. \texttt{WSClean} imager is then used to deconvolve and image the target data. In addition, deconvolution is applied to each subband image(obtained by dividing full bandwidth into 10 subbands with a bandwidth of 82 MHz each).\\
\\
A Multi-Frequency Synthesis (MFS) map was generated with \texttt{WSClean} \cite{2014Offringa}. This map consists of the full bandwidth, which has a central frequency of 1283 MHz in joined deconvolution mode. Auto-masking from \texttt{WSClean} deconvolves each of the 10 subbands with an initial mask of $20\sigma_{\rm rms}$, i.e., 20 times the level of the root-mean-square (RMS) of the image. An artefact-free model of the target is created after the deconvolution process, and it is used for the self-calibration process. For this self-calibration process, tasks from \texttt{Cubical software} are used \cite{2018Kenyon}. Lastly, the final image has a reduced $3\sigma_{\rm rms}$ threshold. At the end, CASA task \texttt{IMFIT} was used to fit a 2D Gaussian to the emission region in the convolved synthesis map, from which the deconvolved major and minor axes were obtained. The fitting region was defined by visual inspection of the intensity distribution rather than a strict signal-to-noise (S/N) threshold.

\section{Results} \label{results}
\begin{figure}[ht]
\centering
\includegraphics[width=1.09\textwidth]{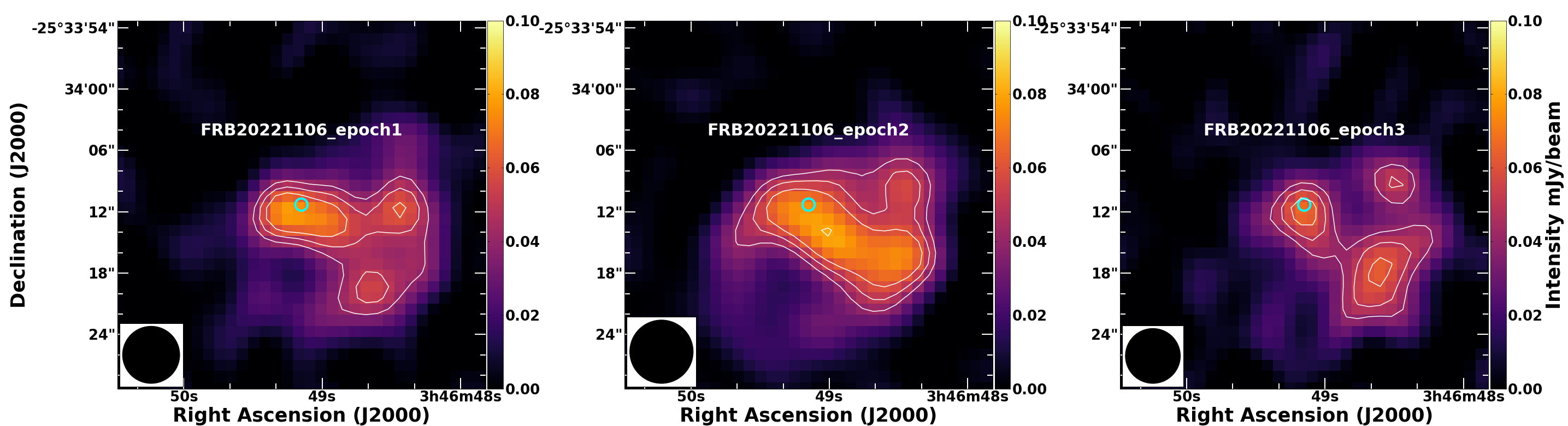}
\caption{\label{FRB3} MeerKAT image of the FRB20221106 position with a $\sim$1$^{\prime\prime}$ localisation uncertainty \cite{Shannon2025} as indicated by the green circle, observed during three different epochs. The black eclipse in the bottom left corner represents the beam size of MeerKAT, and the positional uncertainty of the detected radio source is $\sim$0.3$^{\prime\prime}$, which is smaller than the beam size. The white contours indicate continuum radio emission that coincides with the FRB position, represented at 3, 9, and 12 times the image's RMS.}
\end{figure}
The position of FRB20221106  was observed over three epochs. During these observations, there is consistent detection of radio emissions, and Figure~\ref{FRB3} reveals that FRB20221106 showed three different morphologies, and it also showed three different peak fluxes as indicated in Table~\ref{tab:detected-frb20221106-20181112}.  The small change in peak flux is in the range of $2-3\sigma$, which is not statistically significant. In addition, the change in integrated flux suggests that the source is not static, and shows a possible slow or variable evolution, which can be explained by an afterflow-like scenario. The source also expands as observed in the Maj$\times$Min axis, suggesting structural development. The consistent coincidence of the source with the FRB position, along with the observed flux evolution, suggests that the radio source is associated with the FRB.
\\
This was not the first time flux variability was observed; a similar occurrence was seen in the case of FRB20190520B: A multi-wavelength study of FRB20190520B showed a $\approx20\%$ decrease in flux at 3 GHz over 2 years \cite{Zhang2023}. It was concluded that the variability of FRB20190520B is unlikely due to scintillation, but rather suggests an evolving PRS on yearly timescales.\\

\begin{figure}[ht]
\centering
\includegraphics[width=0.48\textwidth]{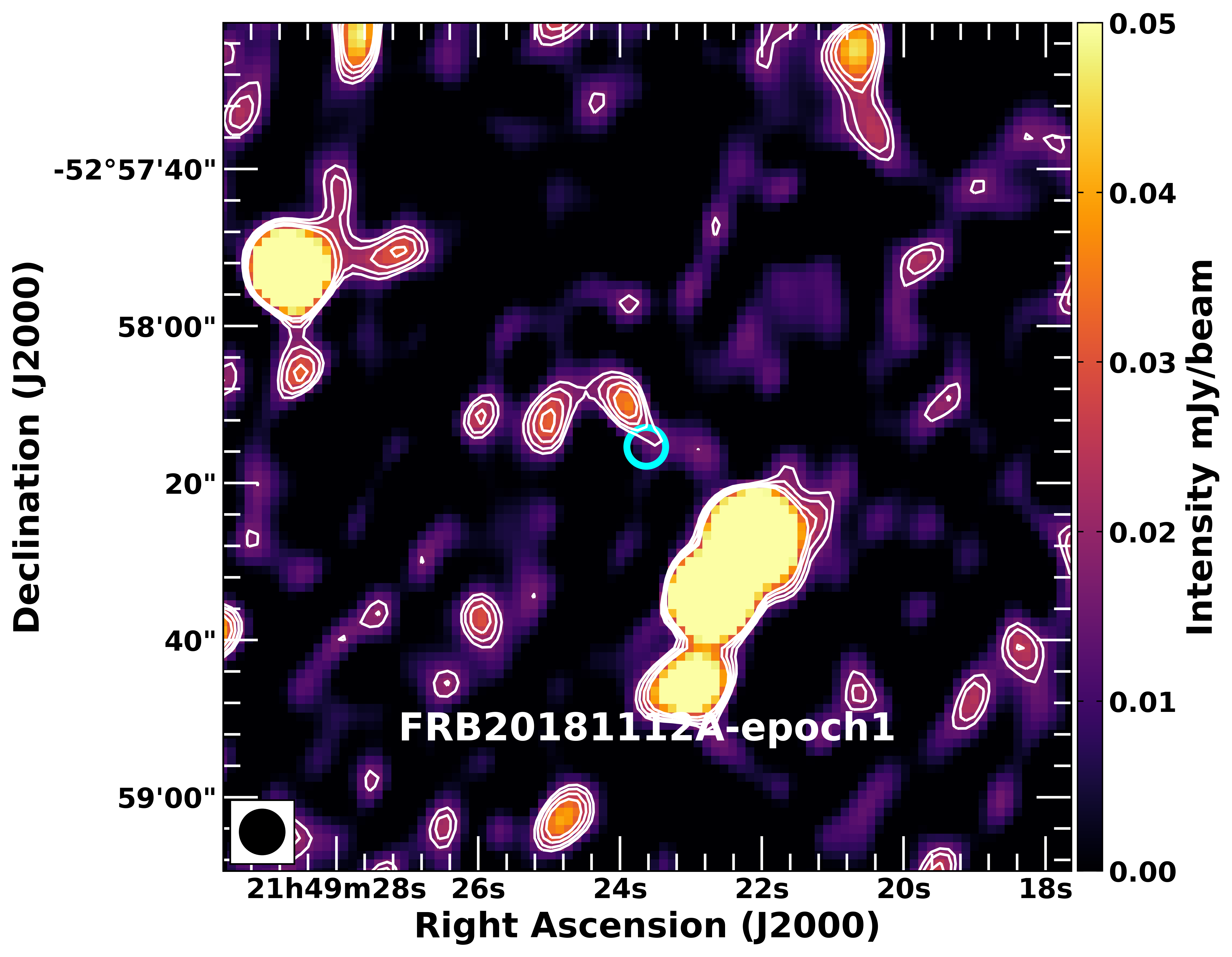}
\includegraphics[width=0.48\textwidth]{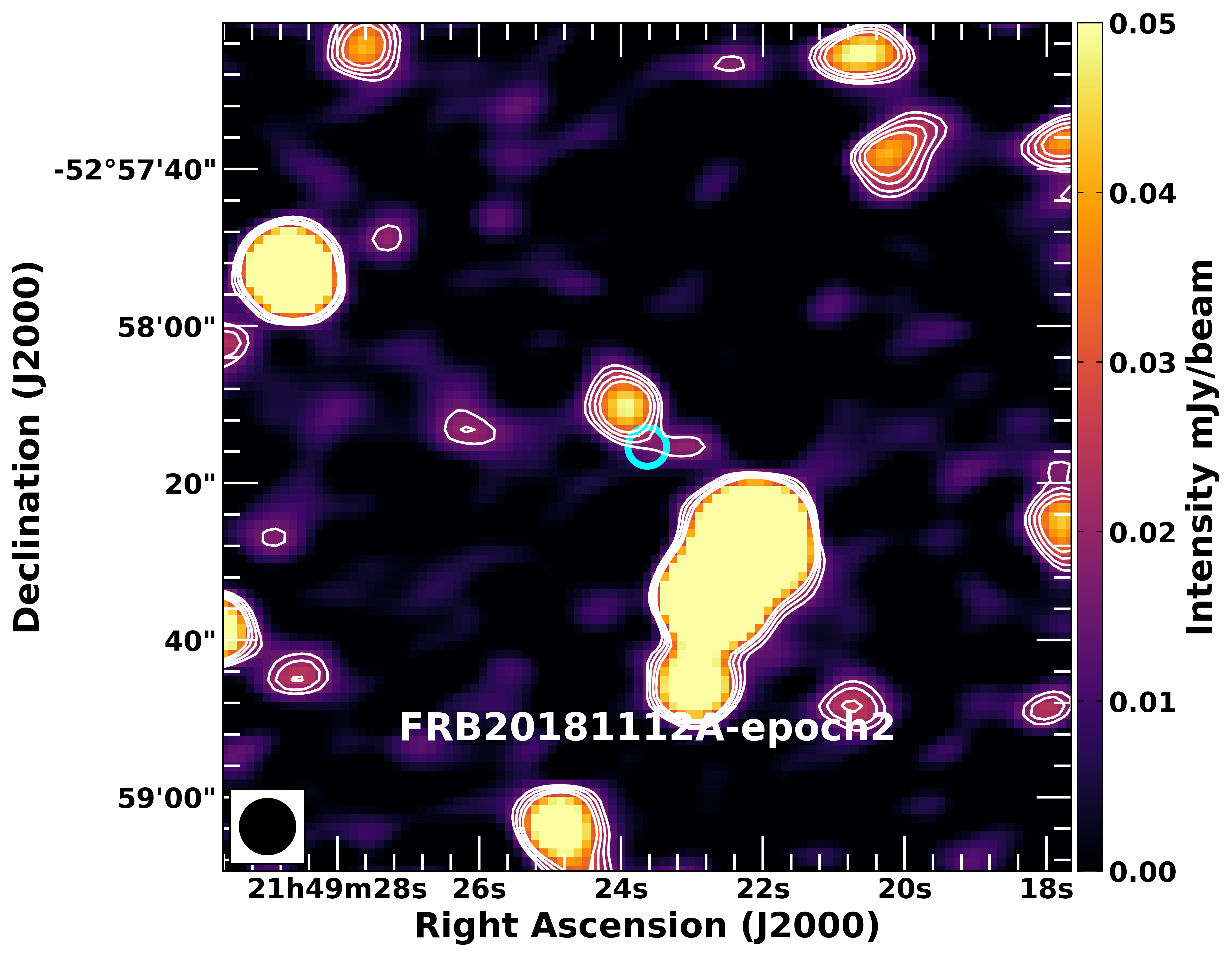}
\caption{\label{fig-onlyone} Same as in Figure 1, but for FRB20181112, observed during two different epochs with $\sim$0.5$^{\prime\prime}$ uncertainty of the FRB position \cite{Prochaska}. The positional uncertainty of the detected radio source is $\sim$0.5$^{\prime\prime}$, again smaller than the beam. Although the emission is faint and extended, the FRB position lies well within the detected source extent.}
\end{figure}

\begin{figure}[htb]
\centering
\includegraphics[width=0.48\textwidth]{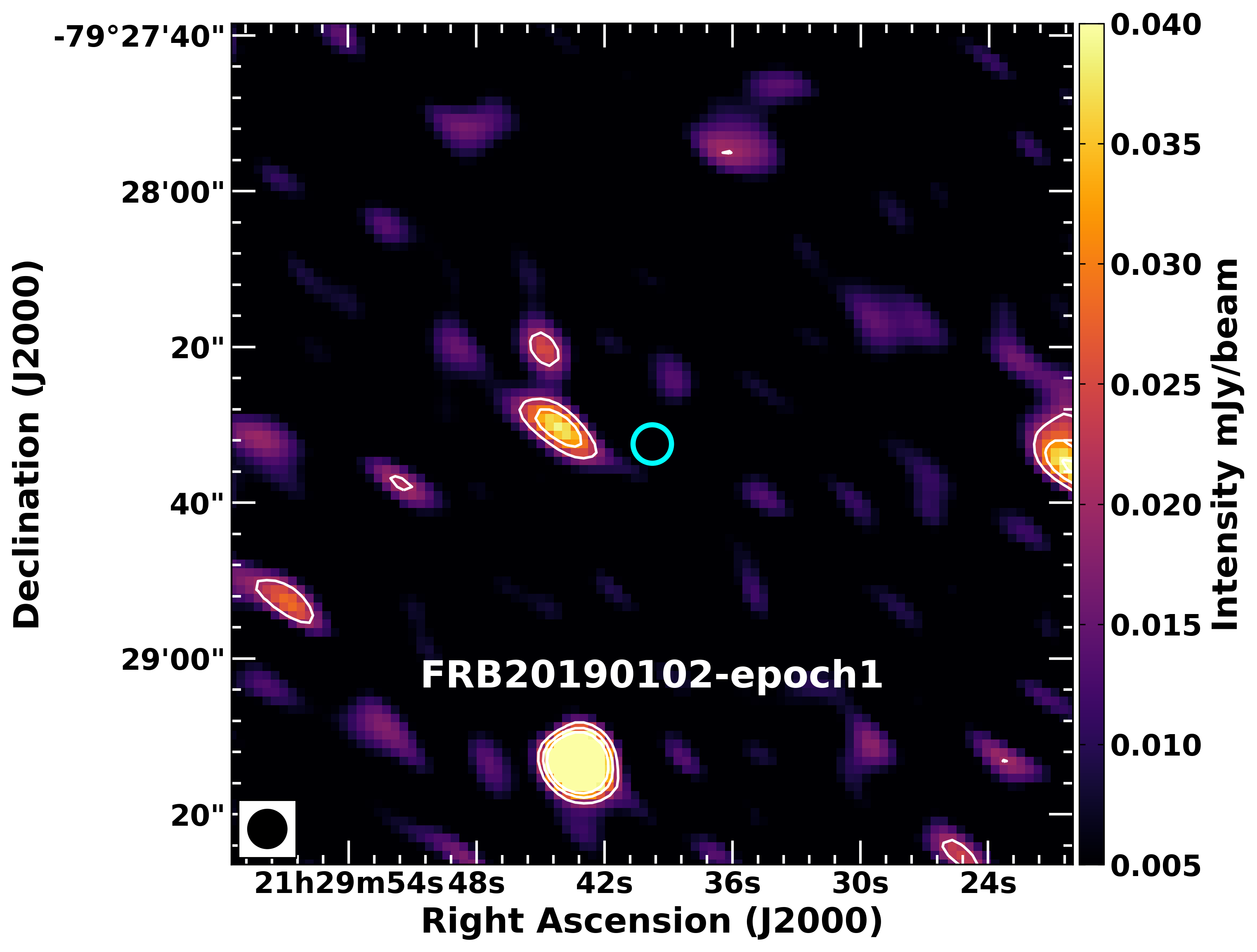}
\includegraphics[width=0.48\textwidth]{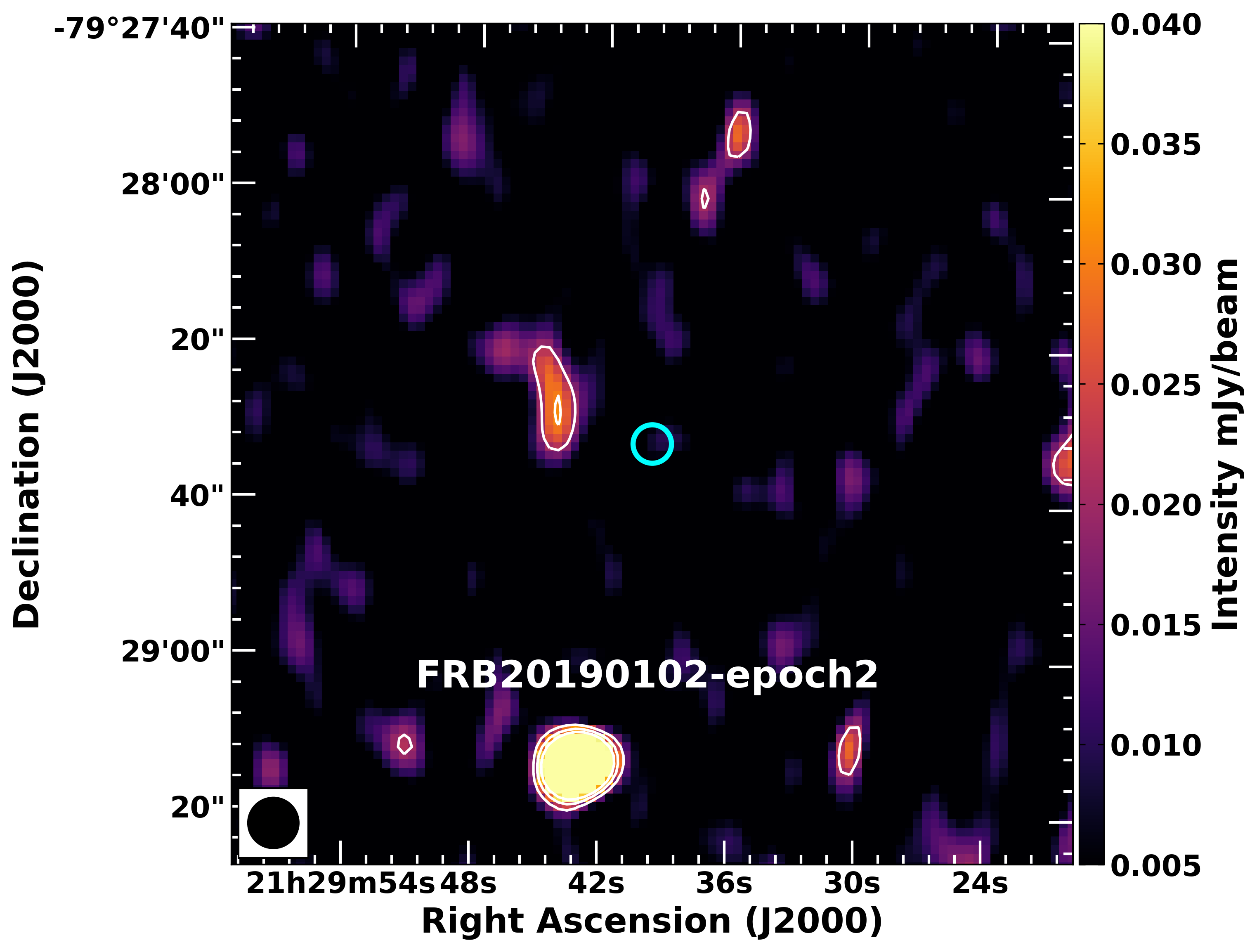}
\caption{\label{fig-sidebyside} Same as Figure 2, but for FRB20190102.}
\end{figure}

\begin{table*}
\centering
\caption{Detected persistent continuum emission from FRB20221106 and FRB20181112.}
\label{tab:detected-frb20221106-20181112}
\resizebox{\textwidth}{!}{%
\begin{tabular}{ccccccccccc}
  \hline
  Source name & Observation date & R.A.(J2000) & Dec.(J2000) & Synthesised beam & RMS & Peak Flux & Maj$\times$Min axis & Pos. Angle & Int. Flux \\
  & & & &  & (mJy beam$^{-1}$) & (mJy beam$^{-1}$) & & & (mJy) \\
  \hline
  FRB20221106 & 21-Nov-2023 & 03:46:49.15 & -25:34:11.3 & 5".649$\times$5".649 & 0.0048 & 0.0656$\pm$ 0.0055 & 20".35$\times$16".38 & 69$^\circ$ & 0.5032 \\
  FRB20221106 & 05-Dec-2023 & 03:46:49.15 & -25:34:11.3 & 6".264$\times$6".264 & 0.0054 & 0.0807$\pm$0.0046 & 34".84$\times$17".61 & 76$^\circ$ & 0.5391 \\
  FRB20221106 & 20-Jan-2024 & 03:46:49.15 & -25:34:11.3 & 5".452$\times$5".452 & 0.0053 & 0.0579$\pm$0.0051 & 38".82$\times$35".64 & 76$^\circ$ & 0.4768 \\
  \hline
  FRB20181112 & 19-Apr-2021 & 21:49:23.63 & -52:58:15.4 & 7".304$\times$7".304 & 0.0047 & 0.0484$\pm$0.0055 & 10".17$\times$7".18 & 44$^\circ$ & 0.0663 \\
  FRB20181112 & 03-Sep-2021 & 21:49:23.63 & -52:58:15.4 & 5".940$\times$5".940 & 0.0054 & 0.0291$\pm$0.0069 & 7".58$\times$3".81 & 38$^\circ$ & 0.0238 \\
  \hline
\end{tabular}%
}
\end{table*}
Another interesting position is that of FRB20181112, which was observed over two epochs (see Figure \ref{fig-onlyone}). While previous studies did not reveal any PRS associated with this FRB \cite{Prochaska}, our observations revealed radio emission in the direction of this FRB. This may indicate that there was low-level persistent emission or maybe afterglow-like activity, which was not detected previously due to sensitivity limits. The measured flux of this FRB appears to decline between the two epochs as shown in Table \ref{tab:detected-frb20221106-20181112}, but the change is within 2$\sigma$, which means it is not statistically significant. Additionally and importantly, a massive halo of gas surrounds the host galaxy of this FRB \cite{Prochaska}. This host galaxy is situated in the southern part of the source, whereby the foreground is in the northern part. It was found that the contributions of this galaxy to DM and scattering are minimal, suggesting weakly magnetised halo gas. If the persistent radio emission is detected during follow-up observations by the e-MERLIN telescope, it is unlikely to be associated with the halo itself and may instead come from the host galaxy or the PRS. \\
\\
\\
The localisation of FRB20221106 by ASKAP has an uncertainty of $\sim$1$^{\prime\prime}$ \cite{Shannon2025}, on the other hand, FRB20181112 was localised to within $\sim$0.5$^{\prime\prime}$ \cite{Prochaska}. Our MeerKAT images have synthesised beams of  $\sim$5--7$^{\prime\prime}$, which makes both FRBs better localised than the beam sizes. The positional uncertainties of the detected radio sources were estimated as $\sigma_{\rm pos} \approx {\rm FWHM}/(2 \times {\rm SNR})$ \cite{Condon_1997}, which gives $\sim$0.2--0.3$^{\prime\prime}$ for FRB20221106 and $\sim$0.4--0.5$^{\prime\prime}$ for FRB20181112. One can see that these uncertainties are significantly smaller than the apparent source sizes (20--38$^{\prime\prime}$). This confirms that the FRB positions coincide with the detected emission. The absolute extent of the sources depend on a rather uncertain source distance, and thus the extended radio emission may either be due to the host galaxy structure, or the blending of multiple components.
\\
Lastly, we did not detect any radio emission from FRB20190102 over two epochs as shown in Figure \ref{fig-sidebyside}. Table~\ref{tab:info2} shows our derived flux upper limits.
\\
\begin{table*}
\centering
\small
 \caption{Observational information of FRB20190102.}
 \label{tab:info2}
 \begin{tabular}{ccccccc}
  \hline
  Source name & Observation & R.A.(J2000) & Dec.(J2000) & Synthesised beam & rms  & Upper limit\\
   & date & &  &  & mJy beam$^{-1}$ & mJy beam$^{-1}$\\
  \hline
   FRB20190102 & 10-Apr-2021 &21:29:39.76 &-79:28:32.5&5".709$\times$ 5".709&0.0054&$< 0.0162$\\
  FRB20190102 &05-Sep-2021 &21:29:39.76 &-79:28:32.5&6".723$\times$ 6".723&0.0055&$< 0.0164$\\
  \hline
    \end{tabular}
    \label{tab:non-detection}
\end{table*}
%\clearpage
\section{Conclusion} \label{conclusion}
As part of an ongoing study targeting 25 FRB positions, we presented results for 3 of them here. Out of the three FRB positions, we have two detections and one non-detection. For FRB20221106, we have positive detections for observations during three epochs. The position of the FRB20221106 coincides with a galaxy; whether the detected radio source originates from the PRS or the galaxy will remain to be seen. To answer this question, a telescope with high angular resolution, such as e-MERLIN, is required. Such a telescope will help us to identify compact PRSs from the observed radio sources. In addition, it will provide the size of the compact PRS if detected, and constrain its spectral shape and variability. In the case of FRB20181112, a telescope with high angular resolution is also needed to elucidate whether the host galaxy is the one providing the radio emissions, or if the emissions come from the PRS. Lastly, for FRB20190102, a flux upper limit was provided. To resolve the detected radio sources, a proposal for a telescope with high angular resolution will be written, which will help to determine if we have detected PRSs with MeerKAT.
%\clearpage
\section*{Acknowledgements}
This paper utilises MeerKAT L-band observational data from 2021, 2022, and 2023 open-time proposals (SCI-20210212-CV-01, SCI-20220822-CV-01, SCI-20230907-CV-01, respectively). The South African Radio Astronomy Observatory, a facility of the National Research Foundation, operates the MeerKAT telescope. The Inter-University Institute for Data Intensive Astronomy (IDIA) visualization lab https://idia.ac.za/citing-idia-ilifu-in-publications/ is used to for this work. IDIA is a partnership of the University of Cape Town, the University of Pretoria, the University of the Western Cape and the South African Radio astronomy Observatory. Thulo Letsele is grateful for financial support from the National Astrophysics and Space Science Programme (NASSP), which is a multi-institutional initiative, funded by DSI-NRF to train South African students in Astrophysics and Space Science.

%\clearpage
%\newpage
%Bibliography
%make sure you fill the .bib properly, e.g. edit "biblio.bib" and add all your references in there.
\bibliographystyle{IEEEtran}
\bibliography{biblio}

\end{document}